\documentclass[preprint]{aastex}

\usepackage{xspace}

\newcommand{\SNR}{HFPK\,334\xspace}
\newcommand{\ROSAT}{\emph{ROSAT}\xspace}
\newcommand{\xmm}{\emph{XMM-Newton}\xspace}
\newcommand{\chan}{\emph{Chandra}\xspace}

\def\HI{\hbox{H\,{\sc i}}}

\def\p0{\phantom{0}}

\shorttitle{HFPK\,334: An unusual SNR in the SMC}
\shortauthors{Crawford et al.}

\begin{document}


\title{HFPK\,334: An unusual Supernova Remnant in the Small Magellanic Cloud}


\author{E.~J.~Crawford and M.~D.~Filipovi\'c}
\affil{University of Western Sydney, Australia}
\email{e.crawford@uws.edu.au}

\and
\author{R. L. McEntaffer, T. Brantseg, K. Heitritter and Q. Roper}
\affil{Department of Physics and Astronomy, University of Iowa, Iowa City, IA 52242}

\and
\author{F.~Haberl}
\affil{Max-Planck-Institut f\"ur extraterrestrische Physik, Giessenbachstra\ss e, 85748 Garching, Germany}

\and
\author{D.~Uro\v sevi\'c}
\affil{Department of Astronomy, Faculty of Mathematics, University of Belgrade, Studentski trg 16, 11000 Belgrade, Serbia}
\affil{Isaac Newton Institute of Chile, Yugoslavia Branch}




\begin{abstract}
We present new Australia Telescope Compact Array (ATCA) radio-continuum and \xmm/\emph{Chandra X-ray Observatory} (CXO) observations of the unusual supernova remnant \SNR in the Small Magellanic Cloud (SMC). The remnant follows a  shell type morphology in the radio-continuum and has a size of $\sim$20~pc at the SMC distance. \textbf{The X-ray morphology is similar, however, }we detect a prominent point source close to the center of the SNR exhibiting a spectrum with a best fit powerlaw \textbf{with a photon index} of $\Gamma = 2.7 \pm 0.5$. \textbf{This central point source is most likely a background object} and cannot be directly associated with the remnant. The high temperature, nonequilibrium conditions in the diffuse region suggest that this gas has been recently shocked and point toward a younger SNR with an age of $\lesssim 1800$ years. With an average radio spectral index of $\alpha=-0.59\pm0.09$ we find that an equipartition magnetic field for the remnant is $\sim$90~$\mu$G, a value typical of younger SNRs in low-density environments. Also, we report detection of scattered radio polarisation across the remnant at 20~cm, with a peak fractional polarisation level of 25$\pm$5\%. 
\end{abstract}



\section{Introduction}

The study of supernova remnants (SNRs) in nearby galaxies is of major interest in order to understand the multi-frequency output of more distant galaxies, and to understand the processes that proceed on local interstellar scales within our own Galaxy. Unfortunately, the distances to many Galactic remnants are uncertain by a factor of $\sim2$ \citep[eg.][]{2009AJ....138.1615J,2009BASI...37...45G,GREEN-CATA}, leading to a factor of $\sim4$ uncertainty in luminosity and of $\sim5.5$ in the calculated energy release of the initiating supernova (SN). At an assumed distance of $\sim60$ kpc \citep{2005MNRAS.357..304H}, the Small Magellanic Cloud (SMC) is one of the prime targets for the astrophysical research of objects, including SNRs. These remnants are located at a known distance, yet close enough to allow a detailed analysis.
 
 \textbf{SNRs reflect a major process in the elemental enrichment of the interstellar medium (ISM). Multiple supernova explosions over space-time generate super-bubbles typically hundreds of parsecs in extent. Both are among the prime drivers controlling the morphology and the evolution of the ISM. Pulsar wind nebulae (PWN) offer further information, as the SNR shell and PWN together provide unique constraints and insights into the ISM. Their properties are therefore crucial to the full understanding of the galactic matter cycle.}

Today, a total number of 24 classified SNRs are known in the SMC \citep[and references therein]{2012A&A...537L...1H,2012A&A...545A.128H,2008A&A...485...63F,2004A&A...421.1031V}. \textbf{This represents the most complete sample of SNRs in any galaxy. There is one confirmed PWN in the SMC, IKT\ 16 \citep{2011A&A...530A.132O}, and at least three other candidate PWNs, \SNR, DEM\,S5 and IKT\ 4. The other 20 SNRs range from the very young 1E0102 at 1400 years old to the very old HFPK\,419 at 50000 years old \citep{2012A&A...537L...1H}, giving an unparalleled insight into the evolution of of SNRs and their environment.
}

\citet{1999A&AS..136...81K} first detected \SNR with R\"ontgensatellit (\ROSAT), and list it as  source 179 in their catalogue. \citet{2000A&AS..142...41H} used advanced data processing and additional available data to extend the \citet{1999A&AS..136...81K} catalogue and noted that \SNR was extended at 13~cm. It was \citet{2008A&A...485...63F} that provided the first conformation that \SNR was an SNR, albeit an unusual SNR with detectable radio and X-ray emission, but no optical emission \citep{2007MNRAS.376.1793P}. They also noted a possible central source, leading them to suggest that it may be a PWN, possibly the first (at the time) detected in the SMC. Here, we present \textbf{new follow-up  radio-continuum observations with the} Australia Telescope Compact Array (ATCA). \textbf{This is in addition to previous higher frequency study of \citet{2008A&A...485...63F}. We also present} new \chan\ X-ray observations, \textbf{together with archival \xmm observations}  of the SMC SNR  \SNR. \textbf{Therefore we present a new insight on \SNR and clarify the nature of the central point source suggested in previous studies \citep{2008A&A...485...63F}.}

\section{Observations}

 \subsection{The ATCA radio-continuum observations and data reduction.}

We observed \SNR\ with the ATCA on 2009 January~5 using the 6C array, and on 2009 February~4 using the EW352 array at wavelengths of 20 and 13~cm $(\nu=1384 \mathrm{~and~} 2367 \mathrm{~MHz})$. The observations were done in an interleaved  mode, totaling $\sim$4~hours of integration over a \textbf{12-hr} period. Source 1934-638 was used for primary calibration and source 0252-712 was used for secondary calibration. The \textsc{miriad} \citep{1995ASPC...77..433S} and \textsc{karma} \citep{1996ASPC..101...80G} software packages were used for data reduction and analysis. Images were formed using \textsc{miriad}'s multi-frequency synthesis algorithm \citep{1994A&AS..108..585S} and natural weighting. They were deconvolved with primary beam correction applied. The same procedure was also used for both {\it Q} and {\it U} Stokes parameter maps. The mean fractional polarisation at 20~cm was calculated using flux density and polarisation:
\begin{equation}
P=\frac{\sqrt{S_{Q}^{2}+S_{U}^{2}}}{S_{I}}\cdot 100\%
\end{equation}
\noindent where $S_{Q}, S_{U}$ and $S_{I}$ are integrated intensities for \textit{Q}, \textit{U} and \textit{I} Stokes parameters. 

\begin{figure}[htbp]
 \centering
\includegraphics[angle=-90,width=\textwidth]{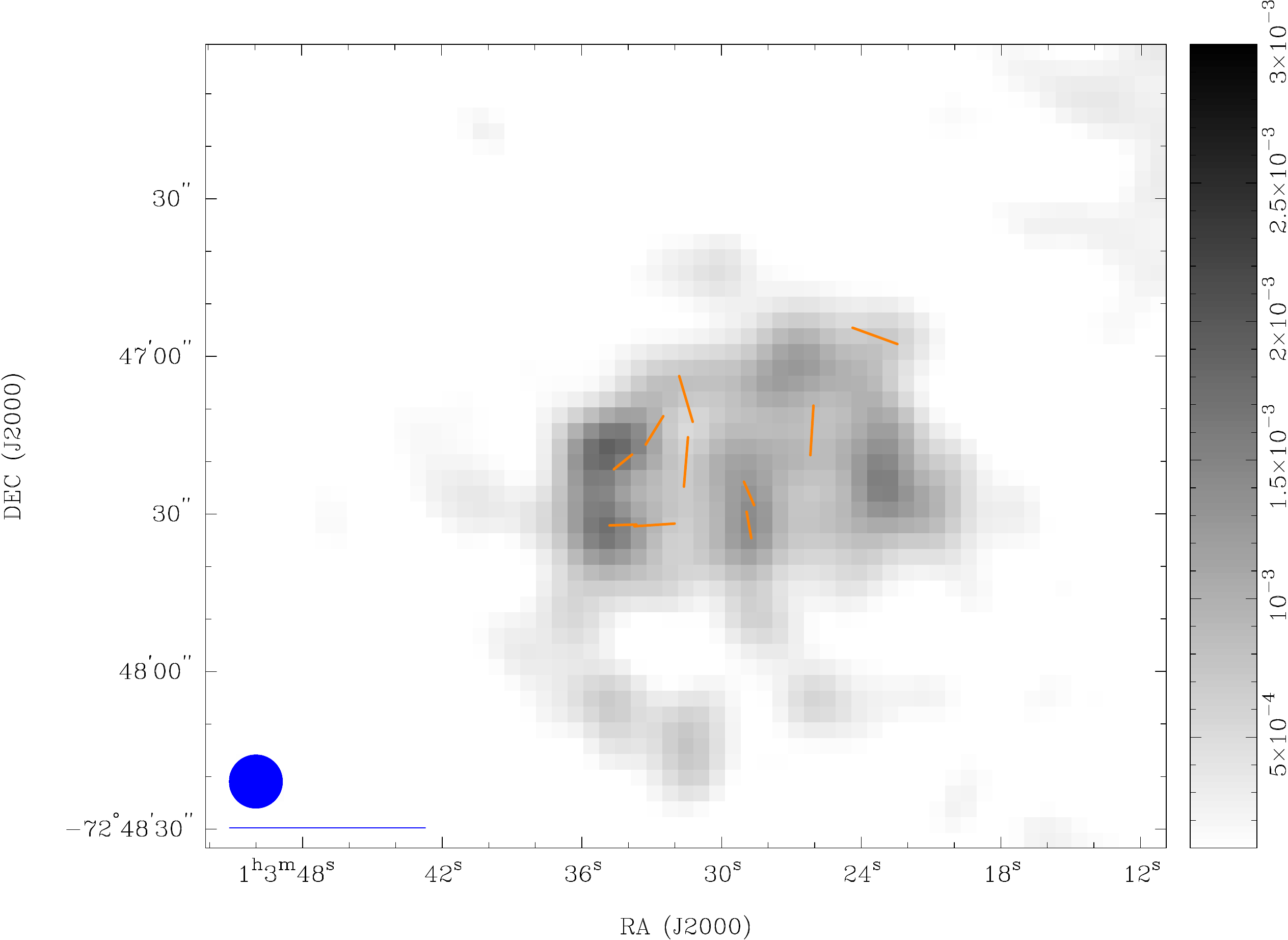}
   \caption{ATCA observations of \SNR\ at 20~cm (1384 MHz) overlaid with fractional polarised intensity. The ellipse in the lower left corner represents the synthesised beam of $10\arcsec$ and the line below the ellipse is a polarisation vector of 100\%. The peak polarisation is estimated to $\sim$25$\pm$5\%. The sidebar quantifies the pixel map in units of Jy/beam. }
           \label{20cmrad}
    \end{figure}

The 20~cm image (Figure~\ref{20cmrad}) has a resolution of $10\arcsec$ and an r.m.s. noise of 0.2~mJy/beam. A matched 13~cm image was produced, with an r.m.s. noise of 0.3~mJy/beam and used in the calculation of the spectral index.

 \subsection{\xmm\ observations and data reduction}
  \label{xmm}

\SNR\ was serendipitously observed during an \xmm\ observation of a candidate supersoft source in the direction of the SMC. The observation (Obs. Id. 0402000101) was performed on 2006 October~3 and further details about instrument setups and data quality can be found in \citet{2008A&A...484..451H}. Using data from this observation, \citet{2008A&A...485...63F} classified the source as a new candidate SNR \textbf{(plerionic)} and presented images \textbf{(their Fig.~5)} and X-ray spectra obtained by the European Photon Imaging Camera (EPIC). However, their spectral analysis of the SNR spectra did not consider the contribution of a point source near the centre of the remnant, which was revealed in follow-up \chan\ observations with superior spatial resolution (see \S\ref{chandra} below).

Therefore, we re-extracted the EPIC-pn \citep{2001A&A...365L..18S} spectra using the \xmm\ Science Analysis System (SAS) version 11.0.0. To obtain the best statistics we selected single- and double-pixel events with quality flag 0 and binned the source spectrum to a minimum of 20 counts per bin. Source counts were extracted from an ellipse with a size of 40\arcsec\ $\times$ 30\arcsec, while the background was selected from a nearby source-free circular region with 60\arcsec\ radius. The net exposure time was 17.5~ks.

 \subsection{The \chan\ X-ray observations and data reduction}
  \label{chandra}

Dedicated X-ray observations of \SNR\ were carried out with the \chan\ X-ray observatory on 2010 December~9 (Obs. Id.~11821) as part of the guaranteed time program available at MPE. The observation was made using ACIS-S and totalled \textbf{$28.55$~ksec}. The level~1 data were reprocessed to level~2 with standard processing procedures in the \chan\ Interactive Analysis of Observations \citep[CIAO, v 4.4;][]{2006SPIE.6270E..60F} software package with current calibration data from the \chan\ Calibration Database (CALDB, v.4.4.2). Good time intervals, charge transfer inefficiency, and time-dependent gain variations were accounted for. The extracted spectra were background-corrected using adjacent regions of the chip that were devoid of emission. \textbf{The effective exposure time after removing bad events is $28.19$~ksec.}

\begin{figure} [htbp]
   \centering
   \includegraphics[width=1\textwidth]{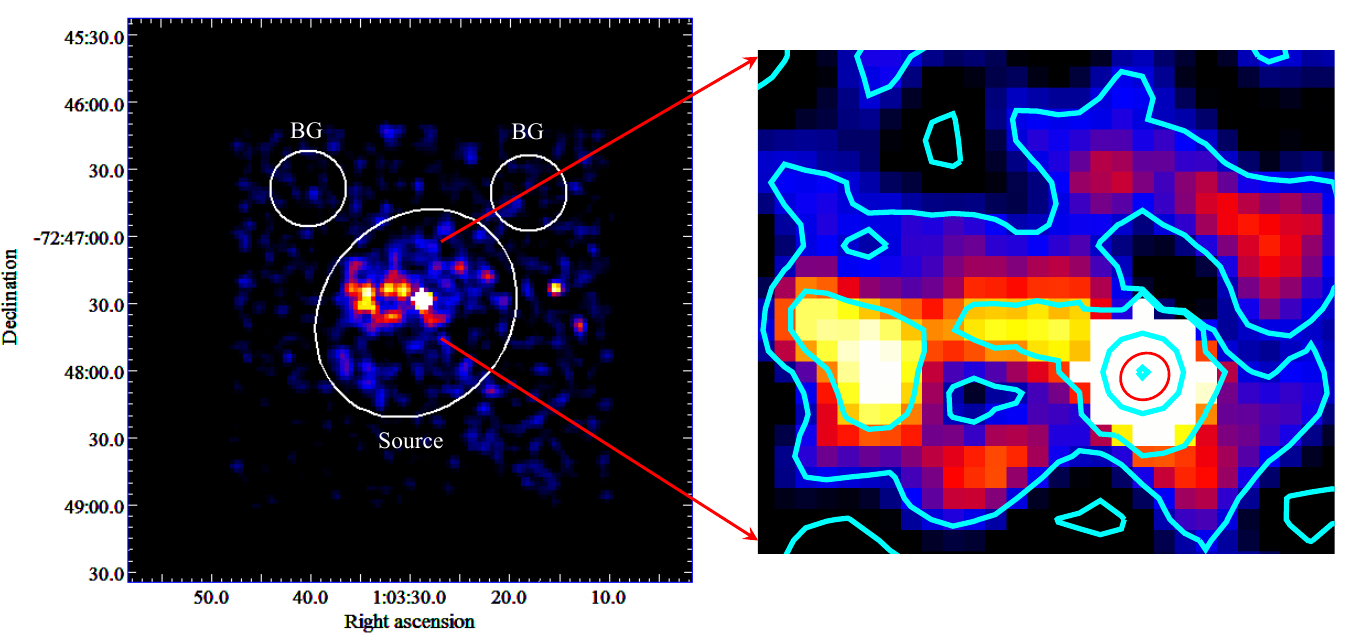}
   \caption{\textit{Left}--Image of the raw data from the \chan\ ACIS S2 CCD.  The data have been binned to 2\arcsec\ pixels and Gaussian-smoothed with a 3-pixel kernel radius to a final scale of 0.2--90 counts/pixel.  The white ellipsoidal region labeled ``Source'' is used for spectral analysis with regions labeled ``BG'' serving as background subtraction regions. \textit{Right}--A close-up of the central emission using the same binning and smoothing parameters.  \textbf{Cyan contours outline the emission at 1.0, 1.8, 2.5, 20, and 40 counts per pixel. The red ellipse is the 6 sigma error ellipse for the point source position, as detected using the CIAO tool \textit{wavdetect}.}}
   \label{raw_data}
\end{figure}

The raw data are shown in Figure~\ref{raw_data}. The image has been binned to $\sim2$\arcsec\ pixels and Gaussian-smoothed over a 3-pixel kernel. The scale ranges from 0.2--90 counts/bin. \textbf{Due to the low average count rate of 0.017 cts/sec, we treat the diffuse emission as a single region.} The red, elliptical region on the right side of Figure~\ref{raw_data} encompasses a point source near the center of the remnant. This region is determined using \textit{wavdetect} in CIAO \citep{2002ApJS..138..185F}. This tool finds sources in the data set by correlating the image with ``Mexican Hat'' wavelet functions. The tool then draws an elliptical region around the detected source out to a specified number of standard deviations. In the present case, the detection scales used are 1 and 2 pixels, and the red region size in Figure~\ref{raw_data} is 3$\sigma$ where $\sigma$ is the uncertainty in the intensity distribution of the detected point source given the point spread function (PSF) of \textit{Chandra}. We use the CIAO tool \textit{mkpsfmap} to determine the expected \chan\ PSF at the location of the point source. The extraction region we use for spectral analysis is larger to ensure all point source photons are included. We use an elliptical region centered on the source\textbf{, at $01^\mathrm{h}03^\mathrm{m}28\fs896$ $-72\degr 47\arcmin 28.35\arcsec$,} and twice the size of the 3$\sigma$ error contour with axes of $4.7\times4.3$ pixels or $2.3\arcsec \times2.1$\arcsec. The remainder of the emission from \SNR\ exists in a diffuse nebula surrounding the point source. The ellipse labelled ``Source'' in Figure~\ref{raw_data} is the extraction region used for this diffuse emission\textbf{, with axes of $42\arcsec\times 49\arcsec$.} The 6$\sigma$ error ellipse used as the extraction region for the point source has been excluded from these data prior to fitting.

Spectra for each of these regions are extracted using the \textit{specextract} tool in CIAO. This tool automatically creates ancillary response files and redistribution matrix files for each region. A background for the diffuse emission is created from the two circular regions shown on the exterior of the remnant.  For the point source the background used is an elliptical annulus surrounding the extraction region with the intent of subtracting any diffuse emission that may be confusing the source spectrum.  The outer bound of this background annulus measures $11.5\times10.6$ pixels (15$\sigma$) with an inner bound consistent with the extraction region. The spectral resolution obtained by \chan\ is $\sim$5--20 ($E/ \Delta E$) over the energy band used, 0.5--8.0~keV.  Data below 0.5~keV and above 8.0~keV are not used, owing to a combination of uncertain ACIS calibration and lack of flux. For the diffuse emission, data above 2.0~keV is consistent with zero in the \chan\ data. The data are binned to at least 20 counts per bin to allow the use of Gaussian statistics.

\section{X-Ray Spectral Fitting}

To analyze the extracted spectra, we use the X-ray spectral fitting software, XSPEC \citep{1996ASPC..101...17A}, and the CIAO modelling and fitting package, Sherpa \citep{2001SPIE.4477...76F}.
 
 \subsection{\chan}

The superior spatial resolution of \chan\ allows us to separate the X-ray spectra of the point source and the extended emission from the SNR. For the point source extraction we attempt a variety of fits that examine a range of possibilities for the nature of the emission. \textbf{These include a powerlaw model to see if the emission is consistent with a pulsar wind nebula (PWN), neutron star, or background AGN, and blackbody and neutron star atmosphere models to test for the presence of a compact central object (CCO).} We apply these models to the \chan\ data since we cannot extract a point source separately from the diffuse emission in the \xmm\ data.

In the diffuse region, if the emission is caused by shock-heating of the ejecta or interstellar medium (ISM), we expect to see prominent emission lines. Depending on the plasma density and the time since it has been shocked, we expect the emitting material to be in either collisional ionization equilibrium (CIE; if the plasma has high density or was shocked a long time ago) or in nonequilibrium ionization (NEI). We use the \textit{xsvnei} model for nonequilibrium conditions \citep{1994ApJ...429..710B,2001ApJ...548..820B,1983ApJS...51..115H,1995ApJ...438L.115L}.  \textbf{This model simulates a single-temperature plasma uniformly shocked at a specific time in the past. Although this assumption is very simplified, as SNR plasmas contain a wide range of temperatures and ionization timescales, our data is of insufficient quality to provide meaningful constraints on the more physically realistic \textit{xsvpshock} or \textit{xsvsedov} models, which account for the range of temperatures and ionization timescales produced by the passage of a plane-parallel (\textit{xsvpshock}) or spherical (\textit{xsvsedov}) model. We therefore treat the plasma temperature and ionization timescale obtained with this model as average values.} We replace the default \textit{xsvnei} line list  with an augmented list developed by Kazik Borkowski that includes more inner shell processes especially for the Fe-L lines \citep{2006ApJ...645.1373B}. For equilibrium conditions we use the \textit{xsvapec} model, which uses an updated version of the ATOMDB code \citep[v2.0.1;][]{2001ApJ...556L..91S,2011nlaw.confC...2F} to model the emission spectrum. We include a second temperature component to these fits if it is determined statistically relevant as inferred from an F-test (probabilites $<0.05$ indicate a statistical improvement given the additional component).  In addition, we investigate the significance of a contribution from a non-thermal component by including the \textit{xssrcut} and$/$or \textit{xspowerlaw} models in the diffuse region \citep{1999ApJ...525..368R,1998ApJ...493..375R}.

To account for interstellar absorption along the line of sight, the above models are convolved with two photoelectric absorption models (\textit{xsphabs}), one of which is held at our Galactic column density along this line of sight, $N_H=2.8\times10^{20}$ cm$^{-2}$ (a high resolution foreground \HI\ map was kindly provided by Erik M\"uller, see also \citet{2003MNRAS.339..105M}), with solar abundances, and the other allowed to vary to account for absorption in the SMC at SMC abundances.  The SMC abundances are taken from \citet{1992ApJ...384..508R} and relative to solar are: He 0.83, C 0.13, N 0.05, O 0.15, Ne 0.19, Mg 0.24, Si 0.28, S 0.21, Ar 0.16, Ca 0.21, Fe 0.20, and Ni 0.40.  Emission line lists in the 0.5--2.0 keV energy range for plasmas with temperatures $kT\sim$0.09--2.0 keV show that the emission is dominated by highly ionized states of C, N, O, Ne, and Fe with contributions from Mg and Si.  The spectral fits begin with all abundances frozen to SMC levels.  A given element is allowed to vary if it significantly improves the fit. Dielectronic recombination rates are taken from \citet{1998A&AS..133..403M} with solar abundances from \citet{2000ApJ...542..914W} and cross-sections from \citet{1992ApJ...400..699B}.

 \subsection{Simultaneous \chan/\xmm\ analysis}

To better constrain our fit parameters we perform simultaneous fitting of \chan\ ACIS and \xmm\ EPIC-pn data in Xspec. The spatial resolution of \xmm\ is insufficient to extract the point source and diffuse emission separately.  However, the additional data may further constrain the physical properties of the source.  Also, flux variations between the different epochs of the \chan\ and \xmm\ data may indicate variability in the central source.  The fit data consist of three inputs: the point source extraction from \chan, the diffuse emission extraction from \chan, and the entire source in the \xmm\ data.  The parent fit model is the same as that found for the \chan\ data, a non-equilibrium ionization model for the diffuse component with an additional powerlaw component for the point source.  Each component is allowed a separate photoelectric absorbing column in the LMC with a global, galactic absorption also applied with a frozen value of $N_H=2.8\times10^{20}$ cm$^{-2}$, as described above [\textit{phabs}$\times$(\textit{phabs}$\times$\textit{vnei}$+$\textit{phabs}$\times$\textit{powerlaw})].  The three data sets are fit simultaneously with the parent model using the following method: the \chan\ point source is fit with the thermal model normalization set to 0; the \chan\ diffuse source is fit with the powerlaw norm set to 0; the \xmm\ data norms for both components are allowed to vary; all thermal models have their abundances, absorbing column, ionization parameter, temperature, and normalization linked; all powerlaw models have their absorbing column, and photon index linked; the \chan\ point source powerlaw norm is frozen to its best fit value while the \xmm\ powerlaw norm is thawed to investigate variability.  We performed various iterations of freezing/thawing various parameters to find the tightest parameter constraints, e.g. thawing the \chan\ point source powerlaw norm, but find no additional limitations.  Furthermore, these additional fit parameters do not result in a lower reduced $\chi^2$.  The final fit with the best statistic and parameter constraints is shown in Table~\ref{table1}.

\begin{deluxetable} {l c c c c c c c c}
\tablewidth{0 pt}
\tablecaption{Best fit parameters for the spectra from different extraction regions}
\tablehead{
	\colhead{Region} &
	\colhead{$\chi^2$/dof} &
	\colhead{$N_{H,PL}$}\tablenotemark{a} &
	\colhead{$N_{H,vnei}$}\tablenotemark{b} &
	\colhead{$\Gamma$} &
	\colhead{$kT$} &
	\colhead{$\tau$}\tablenotemark{c} &
	\colhead{$norm_{PL}$} &
	\colhead{$norm_{vnei}$}	\\
	& & ($10^{22}$cm$^{-2}$) & ($10^{22}$cm$^{-2}$) & & (keV) & (10$^9$ cm$^{-3}$s) & 10$^{-5}A$\tablenotemark{d} & 10$^{-5}B$\tablenotemark{e}
}
\startdata
Point source & 13/19 & 0.8$\pm$0.4 & \nodata & 2.7$\pm$0.5 & \nodata & \nodata & 5$\pm$2 & \nodata \\
Diffuse emission & 18.0/19  & \nodata & $<$0.5& \nodata & 1$^{+3}_{-1}$ & 6$\pm$3 & \nodata & 5$^{+21}_{-2}$\\
Combined\tablenotemark{f} & 84/89 & 0.8$\pm$0.1 & 0.1$^{+0.3}_{-0.1}$ & 2.8$\pm$0.2 & 1.3$^{+1}_{-0.8}$ & $7^{+5}_{-2}$ & 8$\pm$1\tablenotemark{g} & 6$^{+0.2}_{-2}$\\
\enddata
\tablenotetext{a}{Absorption column in the SMC along the line of sight to the point source.}
\tablenotetext{b}{Absorption column in the SMC along the line of sight to the diffuse emission.}
\tablenotetext{c}{Ionization timescale, defined as $n_e t$, where $t$ is the time since the plasma was shocked.}
\tablenotetext{d}{Normalization parameter. $A$ = 1 photon keV$^{-1}$ cm$^{-2}$ s$^{-1}$ at 1 keV.}
\tablenotetext{e}{Normalization parameter. $B=[10^{-14}/(4 \pi D^2)] \int n_e n_H dV$, where $D$ is the distance to the SMC (60 kpc) and the integral is the volume emission measure.}
\tablenotetext{f}{This is a parallel fit with \xmm, \chan\ point source, and \chan\ diffuse emission data sets.}
\tablenotetext{g}{Normalization parameter for the \xmm\ data.  The \chan\ data are set to the best fit normalization.}
\tablecomments{Quoted errors are 90\% confidence intervals for the parameter in question.}
\label{table1}
\end{deluxetable}

\section{Discussion}

 \subsection{Radio}

\SNR\ has a clumpy appearance, with a knot of emission at the centre, which lead \citet{2008A&A...485...63F} to suggest that it was likely to contain a PWN, centered at $01^\mathrm{h}03^\mathrm{m}29\fs5$  $-72\degr 47\arcmin 20\arcsec$ with the enclosing remnants extent of $70\arcsec\times$40\arcsec\ (20$\times$12~pc) at PA=--70\degr. Using the flux density measurements and images of \citet{2008A&A...485...63F} along with our new measurements at 20~cm and 13~cm of 26.9$\pm1.3$~mJy and 18.9$\pm1.5$~mJy respectively, we estimate a spectral index $\alpha=-0.59\pm0.09$ where {$S_\nu\propto\nu^\alpha$}. This estimate includes emission from the entire remnant and central object. 

\begin{figure}[htbp]
 \centering
\includegraphics[width=1\textwidth]{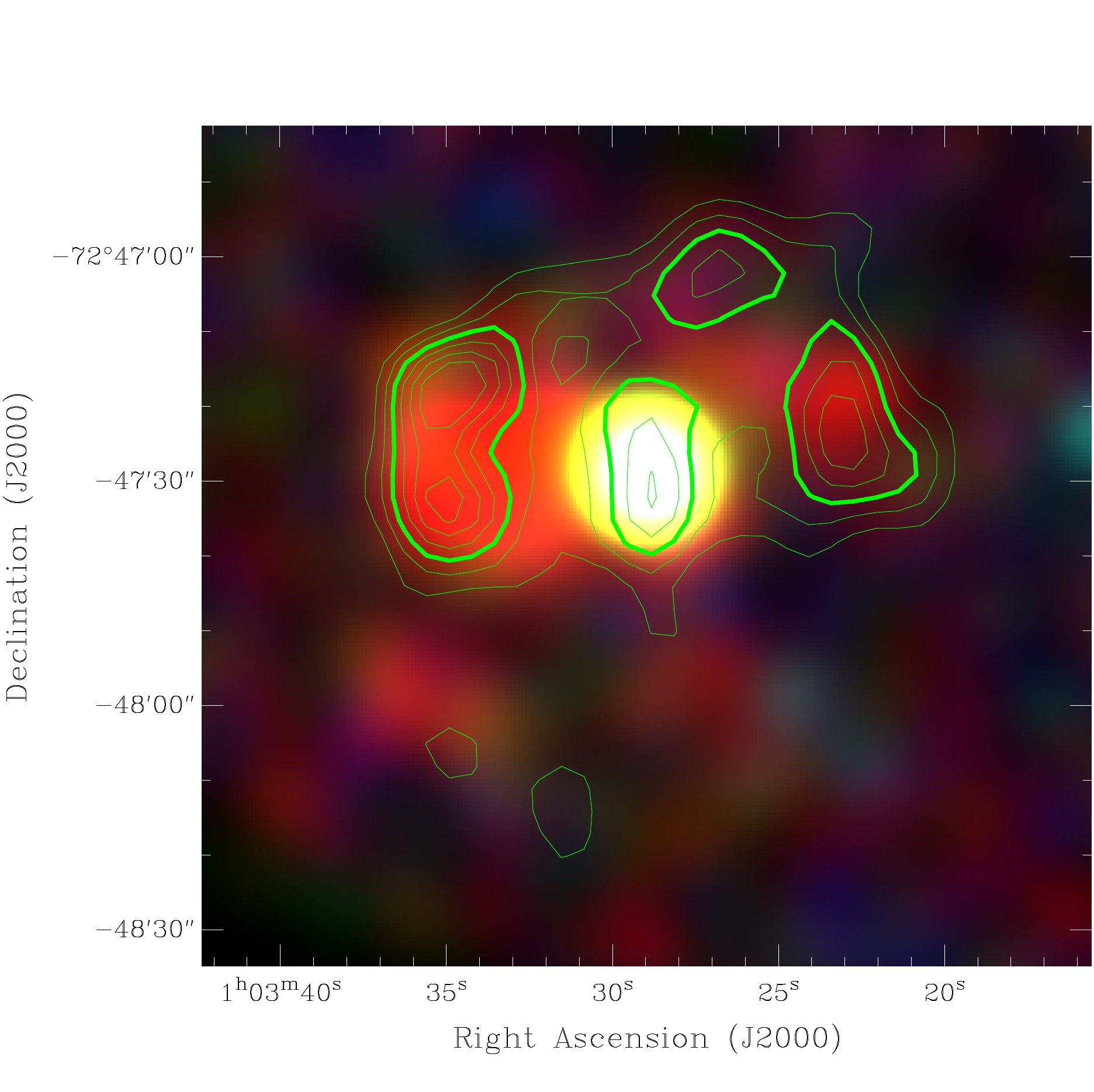}
   \caption{\chan\ three colour composite (red: 0.3--1.0~keV (soft), green: 1.0--2.0~keV (medium), blue: 2.0--6.0~keV (hard)) smoothed with a Gaussian width to match 20~cm radio image. The radio contours
are 0.6--1.6~mJy/beam in 0.2~mJy/beam steps.}
           \label{xrayv20}
    \end{figure}

We also created an image from just the longest ATCA baselines (i.e. those to ATCA antenna 6), which shows no indication of a central point source, to a $3\sigma$ detection limit of $\sim0.3$~mJy. However, the extended radio emission aligns well with the diffuse X-ray emission (Figure~\ref{xrayv20}). 

Linear polarisation images were also formed, as shown in Figure~\ref{20cmrad}. The polarised emission is unordered, with a maximum of 25$\pm$5\%, and indicates a random magnetic field. We note that this order of polarisation from \SNR\ is relatively high when compared to other SNRs in the MCs\textbf{ for which the typical range is 0--20\%%
\citep{2008SerAJ.177...61C,2008SerAJ.176...59C,2010A&A...518A..35C,
2009SerAJ.179...55C,
2010SerAJ.181...43B,2012SerAJ.184...69B,2012RMxAA..48...41B,2012SerAJ.185...25B,2012MNRAS.420.2588B,2013MNRAS.432.2177B,2014MNRAS.440.3220B,2014MNRAS.439.1110B,
2012A&A...539A..15G,
2012A&A...540A..25D,2014arXiv1404.3823D,
2012A&A...543A.154H,
2012A&A...546A.109M,
2013A&A...549A..99K}.}


With a radio surface brightness of $0.36\times10^{-20}$ W~m$^{-2}$~Hz$^{-1}$~sr$^{-1}$, and a diameter of $\sim$20~pc, the position of \SNR on the surface brightness-diameter diagram of \citet[][their Figure~6]{2004A&A...427..525B}, \textbf{leads us to infer an explosion energy to be in the order of $2\times10^{51}$ ergs}. We calculate an equipartition magnetic field of $\sim 90~\mu$G \citep{2012ApJ...746...79A} 
\textbf{which is high for an MC SNR (Bozzetto et al. in prep). 
Assuming a strong shock passing  through the ISM one can expect magnetic field of up to $\sim20~\mu$G 
(the SMC magnetic field is  $\sim3 \mu$G \citet{2008ApJ...688.1029M}). Another mechanism, so called amplification of the magnetic field (added to simple compression by the shock) is therefore necessary to explain such a high magnetic field of $90~\mu$G. The amplification of magnetic field is process driven by very fast shocks of young SNRs. Because of this, a spectral index of $-0.59$, the location in the surface brightness-diameter diagram, and moderate amplification (strong amplification would lead to a field of a few hundred $\mu$G \citep{2012APh....35..300T}) of magnetic field suggesting that this SNR is on somewhat younger age i.e. $<$5000~yr.}


 \subsection{X-Ray}

\begin{figure} [htbp]
   \centering
   \includegraphics[width=1\textwidth]{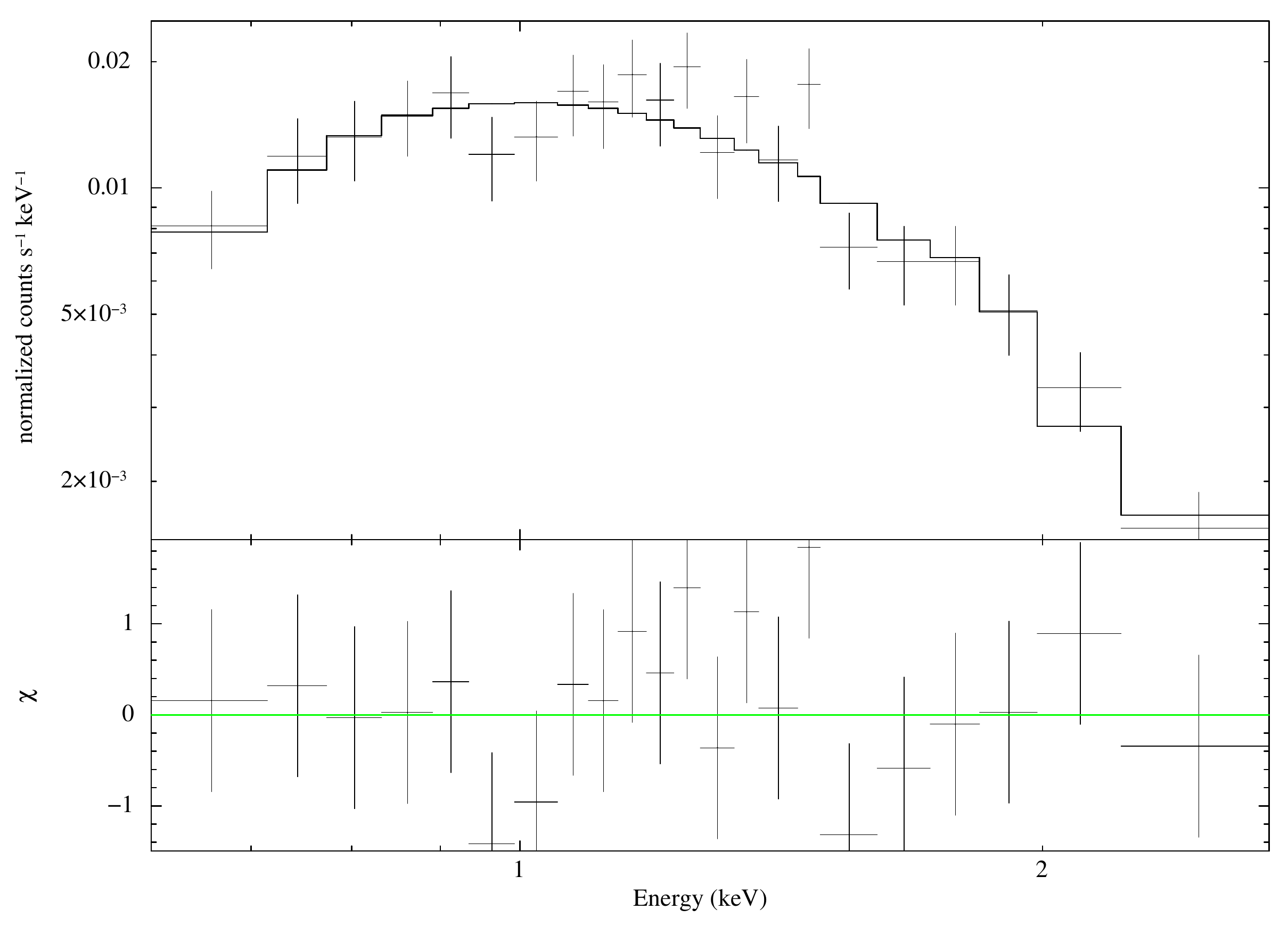}
   \caption{\chan\ spectrum extracted from the point source region with the best fit powerlaw model overlaid.}
   \label{point_spec}
\end{figure}

The best fit parameters for the two \chan\ regions are given in Table \ref{table1}. The point source region is best fit by a powerlaw with photon index $\Gamma=2.7\pm0.5$. The fit is shown overlaid on the data in Figure~\ref{point_spec}. Attempts at fitting the point source spectrum with other possible models\textbf{, such as blackbody and neutron star atmosphere to test for  a leftover compact object,}  were generally reasonable but the absorbing column is poorly constrained. A neutron star atmosphere model, \textit{nsa} in Xspec, resulted in a fit statistic of $\chi^2/dof=15.3/19$ and an effective temperature of $\log{T}=6.5$, but the absorbing column was low, with an upper limit of $3\times10^{20}$ cm$^{-2}$. 

We also model the spectrum using the \textit{bbody} model resulting in a fit statistic of $\chi^2/dof=17.5/19$ and temperature of $kT\sim0.4$~keV.  Again, the absorbing column is ill-constrained, with an upper limit of $1.6\times10^{21}$ cm$^{-2}$. Using the energy band common to both \chan\ and \xmm, 0.5--4.0~keV, we can calculate the fluxes and luminosities for these models. The flux values are consistent and range from $7.5\times10^{-14}$ erg cm$^{-2}$ s$^{-1}$ to $8.7\times10^{-14}$ erg cm$^{-2}$ s$^{-1}$ for the various models resulting in luminosities from $3.3\times10^{34}$ erg s$^{-1}$ to $3.8\times10^{34}$ erg s$^{-1}$ assuming a 60~kpc distance to the object.

The X-ray position of the central point source is $01^\mathrm{h}03^\mathrm{m}28\fs896$ $-72\degr 47\arcmin 28.35\arcsec$. We find a faint (I$_{mag}$=20.849 and V$_{mag}$=21.628) object in OGLE survey \citep{1998AcA....48..147U} at a distance of $\sim$1.52\arcsec. As the 2$\sigma$ error of our X-ray positional estimate is in order of 1.4\arcsec\ and OGLE positional error is $\sim$0.2\arcsec\ we claim that this is fully consistent with an association (according to the cumulative Rayleigh distribution the probability for the counterpart to be found within 1.5\arcsec\ is 88\%).
\textbf{We calculate the optical to X-ray flux ratio to be 0.91, indicating background object, a foreground star would have a ratio less than $-1$ and a foreground neutron star would have a ratio greater than 4 \citep{1988ApJ...326..680M,2004AdSpR..33..638H,2013A&A...558A...3S}.}

\begin{figure} [htbp]
   \centering
   \includegraphics[width=1\textwidth]{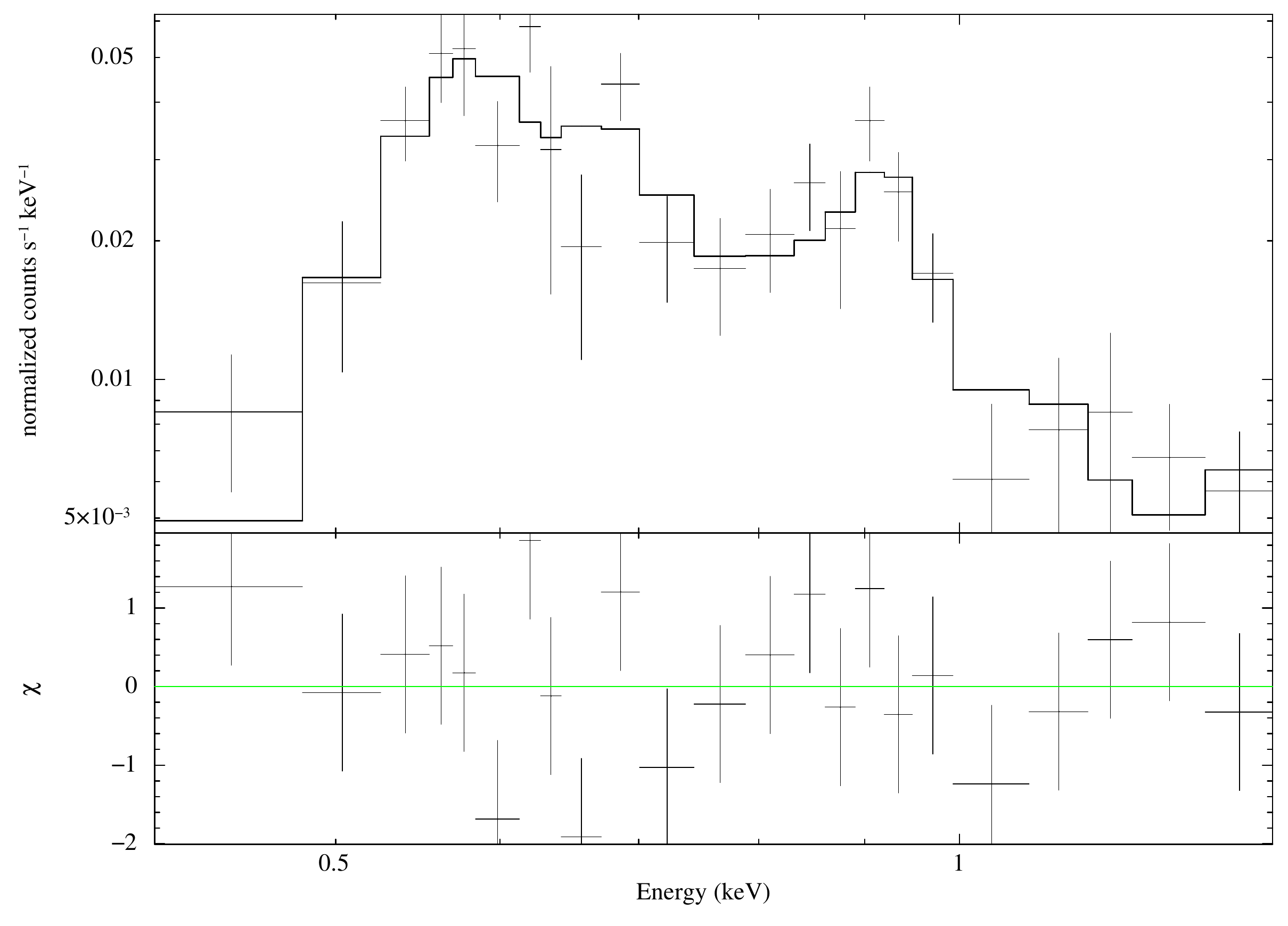}
   \caption{\chan\ spectrum extracted from the diffuse emission region with the best fit thermal model overlaid.}
   \label{diffuse_spec}
\end{figure}

The diffuse emission region cannot be described by a powerlaw and is best fit with a nonequilibrium thermal plasma at a temperature of $kT=1^{+3}_{-1}$~keV. This fit is shown in Figure~\ref{diffuse_spec}. The fit could not be improved with variable abundance of any element or combinations thereof.

\begin{figure} [htbp]
   \centering
   \includegraphics[angle=-90,scale=0.6]{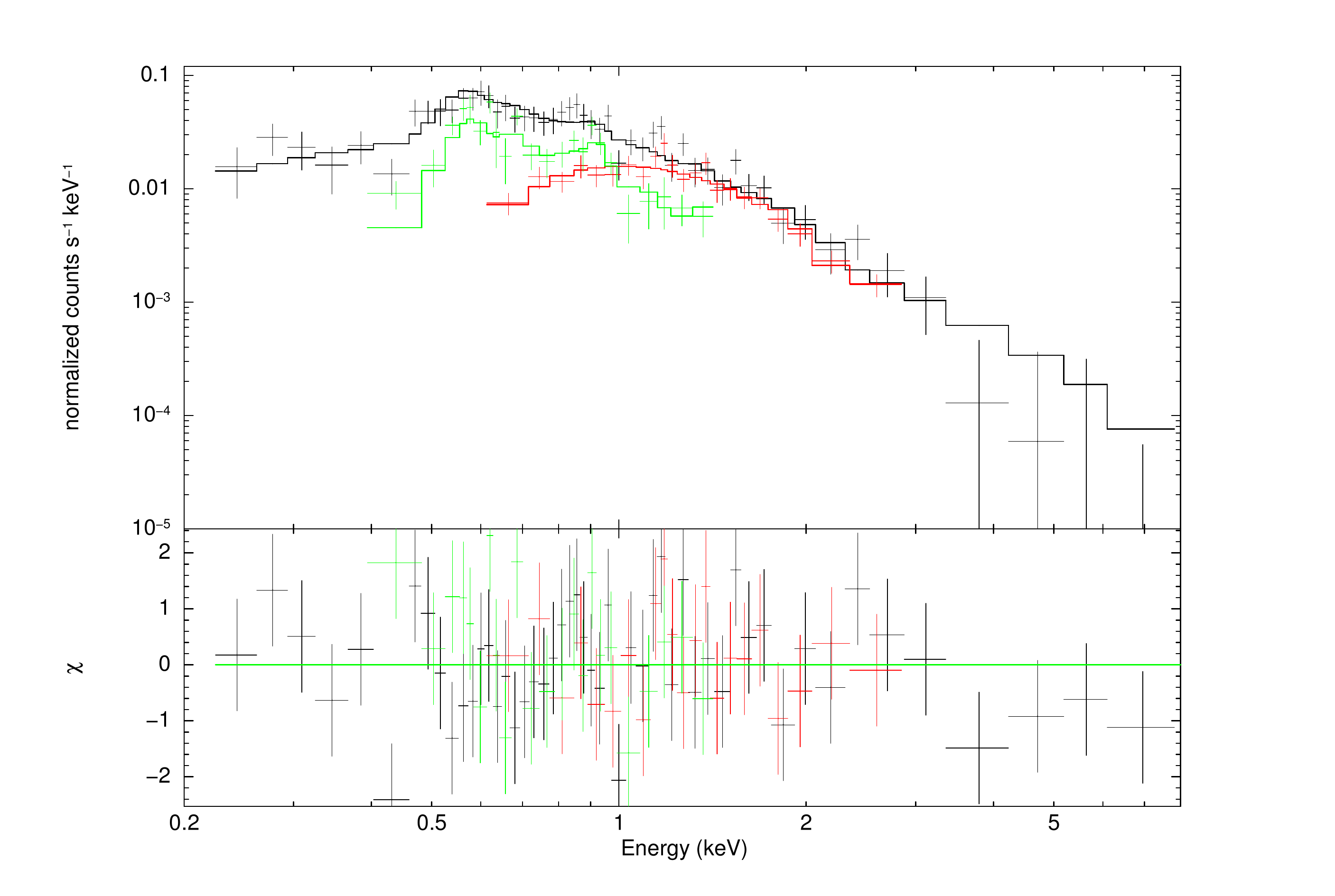}
   \caption{Co-fit \chan\ (point source -- red, diffuse emission -- green) and \xmm\ (black) spectra with best fit model overlaid.}
   \label{chandraxmm}
\end{figure}

The results of the combined fit are also given in Table \ref{table1}. In addition to the powerlaw component for the point source, a nonequilibrium plasma is again found as the best explanation for the diffuse emission. This thermal component has a temperature comparable to that found when considering \chan\ data alone. The additional \xmm\ data allow for tighter constraints on the temperature and ionization parameter for the thermal plasma, and normalizations for both components. \textbf{However, the combined fit is still not improved by thawing any elemental abundances.} The best fits to these data are shown in Figure~\ref{chandraxmm}.

The spectral index of the high energy emission, $\alpha=2.7\pm0.5$, is high, but given the large error bars, barely consistent with the range expected if the emission was arising from a PWN. Given a lack of \textbf{plerionic} remnants in the SMC we can use the LMC for comparison. PWN containing remnants in the LMC typically have a lower index: $\alpha=1.0\pm0.2$ in SNR 0453-68.5 \citep{2012ApJ...756...17M}; $\alpha=0.4-1.4$ for a series of annular regions encompassing the PWN of SNR~0540-69.3 \citep{2007ApJ...662..988P}; $\alpha=1.2\pm0.3$ in N\,206 (SNR~0532-71.0) \citep{2005ApJ...628..704W}; $\alpha=0.57^{+0.05}_{-0.06}$ for DEM\,L241 \citep{2006A&A...450..585B}; and $\alpha\sim1.2-1.6$ over several regions from N\,157B \citep{2006ApJ...651..237C}. \textbf{At SMC distance the calculated luminosity would be consistent with that expected expected from a PWN}. Even though a PWN fits in nicely with the picture of a young SNR and the powerlaw fit was statistically best, the photon index is more compatible with an AGN interpretation, which typically have typical $\Gamma \sim 1.5 - 2.1$ \citep{2010A&A...512A..58I}.  

\textbf{As shown in Table~\ref{table1}, the thermal and powerlaw components of the combined fits have different column densities. We expect that if the point source were associated with the surrounding SNR, the two components would have consistent absorbing columns. The significantly higher column density of the point source implies that it is an unrelated background source.}

The powerlaw fit was reasonable even though we allowed the normalization to vary between the \xmm\ and \chan\ data. We did not detect significant flux variability. The best fit norm for the \xmm\ data is higher than that for the \chan\ data but just consistent with one another at the limits of the 90\% confidence interval. Higher significance data are required to properly constrain the flux over multiple epochs to determine if flux variability is present.

The high temperature, nonequilibrium conditions in the diffuse region suggest that this gas has been recently shocked and point toward a younger SNR. Calculations of shock velocity and density support this conclusion. Solving the Rankine-Hugoniot relations in the strong shock case \textbf{for a monatomic plasma $(\gamma = 5/3)$},  gives the post-shock temperature as a function of shock velocity, $kT=(3/16) \mu m_{p} v^2$. \textbf{We assume that the plasma is fully ionized, so $\mu=0.6$ and $n_e=1.2 n_H$} This equation makes the implicit assumption that $T_e\sim T_{ion}$. This is not  necessarily the case for shock-heated plasmas \citep{2007ApJ...654L..69G}, so our calculated velocities may be taken as a lower bound. The calculated velocity at the best fit temperature is $\sim1100$~km/sec, significantly decelerated from the initial blast wave velocity ($>5000$ km/s), but still very fast. It does not appear that the diffuse emission is from particularly dense material even though it is probably dominated by shock-heated ISM as suggested by the SMC ISM abundances in the fit. The density is found using the \textit{norm} parameter for the diffuse component, where $norm=[10^{-14}/(4 \pi D^2)] \int n_e n_H dV$. The integral contains the emission measure for the plasma which is dependent on the density and total emitting volume. A value of 60 kpc is used for $D$ and we assume that $n_e=1.2n_H$. We use the volume of a sphere with a radius equal to the average of the major and minor axes of the elliptical region used for extraction. We also include a filling factor, $f$, such that $V=4/3 \pi R^3 f$. The resulting electron density is $0.09/\sqrt{f}$ cm$^{-3}$ supporting a low-density environment. Given that the filling factor is always less than 1, and most likely for this remnant much less than 1, this density is a lower limit and most likely larger. The low-density may explain the high temperature, nonequilibrium conditions and a conclusion of a young SNR may be premature. However, we can estimate the age of the remnant from the $\tau$ parameter since $\tau=n_et$, where $t$ is this timescale. This calculation yields a shock time of $1800\sqrt{f}$ years, thus supporting a lower age for this object. Finally, it is quite possible that this diffuse emission cannot be explained using a single set of plasma conditions. There is a need for higher signal-to-noise data to enable more detailed spatially resolved spectroscopy.

\section{Conclusions}
 
\SNR is a young \textbf{shell-type} radio/X-Ray SNR with no optical or IR counterpart. The most striking feature of this SNR is the bright central object seen only in our X-ray observations. We argue that this central object with the best fit powerlaw $\Gamma = 2.7 \pm 0.5$, could not be definitely associated with the remnant, as either a pulsar or compact central object. \textbf{Therefore  we propose that the central point source is a background object}. The remnant itself appears as quite young, $<1800$ years, and 
our estimates of the remnant magnetic field ($\sim$90~$\mu$G) also favours younger age. The somewhat higher temperature and nonequilibrium conditions in the diffuse region suggest that this gas has been recently shocked. We report detections of scattered regions showing moderately high orders of polarisation at 20~cm, with a peak value of $\sim$25$\pm$5\%, indicating the magnetic field is unordered.

\acknowledgments
The Australia Telescope Compact Array is part of the Australia Telescope National Facility which is funded by the Commonwealth of Australia for operation as a National Facility managed by CSIRO. The scientific results reported in this article are based on observations made by the \emph{Chandra X-ray Observatory} (CXO). Based on observations obtained with \xmm, an ESA science mission with instruments and contributions directly funded by ESA Member States and NASA. This research is supported by the Ministry of Education, Science and Technological Development of the Republic of Serbia through project No.~176005.

{\it Facilities:} \facility{ATCA}, \facility{CXO}, \facility{\xmm}.

\bibliographystyle{apj}
\bibliography{references}

\begin{thebibliography}{68}
\expandafter\ifx\csname natexlab\endcsname\relax\def\natexlab#1{#1}\fi

\bibitem[{{Arbutina} {et~al.}(2012){Arbutina}, {Uro{\v s}evi{\'c}},
  {Andjeli{\'c}}, {Pavlovi{\'c}}, \& {Vukoti{\'c}}}]{2012ApJ...746...79A}
{Arbutina}, B., {Uro{\v s}evi{\'c}}, D., {Andjeli{\'c}}, M.~M., {Pavlovi{\'c}},
  M.~Z., \& {Vukoti{\'c}}, B. 2012, \apj, 746, 79

\bibitem[{{Arnaud}(1996)}]{1996ASPC..101...17A}
{Arnaud}, K.~A. 1996, in Astronomical Society of the Pacific Conference Series,
  Vol. 101, Astronomical Data Analysis Software and Systems V, ed. G.~H.
  {Jacoby} \& J.~{Barnes}, 17

\bibitem[{{Badenes} {et~al.}(2006){Badenes}, {Borkowski}, {Hughes}, {Hwang}, \&
  {Bravo}}]{2006ApJ...645.1373B}
{Badenes}, C., {Borkowski}, K.~J., {Hughes}, J.~P., {Hwang}, U., \& {Bravo}, E.
  2006, \apj, 645, 1373

\bibitem[{{Balucinska-Church} \& {McCammon}(1992)}]{1992ApJ...400..699B}
{Balucinska-Church}, M., \& {McCammon}, D. 1992, \apj, 400, 699

\bibitem[{{Bamba} {et~al.}(2006){Bamba}, {Ueno}, {Nakajima}, {Mori}, \&
  {Koyama}}]{2006A&A...450..585B}
{Bamba}, A., {Ueno}, M., {Nakajima}, H., {Mori}, K., \& {Koyama}, K. 2006,
  \aap, 450, 585

\bibitem[{{Berezhko} \& {V{\"o}lk}(2004)}]{2004A&A...427..525B}
{Berezhko}, E.~G., \& {V{\"o}lk}, H.~J. 2004, \aap, 427, 525

\bibitem[{{Borkowski} {et~al.}(2001){Borkowski}, {Lyerly}, \&
  {Reynolds}}]{2001ApJ...548..820B}
{Borkowski}, K.~J., {Lyerly}, W.~J., \& {Reynolds}, S.~P. 2001, \apj, 548, 820

\bibitem[{{Borkowski} {et~al.}(1994){Borkowski}, {Sarazin}, \&
  {Blondin}}]{1994ApJ...429..710B}
{Borkowski}, K.~J., {Sarazin}, C.~L., \& {Blondin}, J.~M. 1994, \apj, 429, 710

\bibitem[{{Bozzetto} {et~al.}(2010){Bozzetto}, {Filipovic}, {Crawford},
  {Bojicic}, {Payne}, {Medik}, {Wardlaw}, \& {de Horta}}]{2010SerAJ.181...43B}
{Bozzetto}, L.~M., {Filipovic}, M.~D., {Crawford}, E.~J., {et~al.} 2010,
  Serbian Astronomical Journal, 181, 43

\bibitem[{{Bozzetto} {et~al.}(2012{\natexlab{a}}){Bozzetto}, {Filipovic},
  {Crawford}, {De Horta}, \& {Stupar}}]{2012SerAJ.184...69B}
{Bozzetto}, L.~M., {Filipovic}, M.~D., {Crawford}, E.~J., {De Horta}, A.~Y., \&
  {Stupar}, M. 2012{\natexlab{a}}, Serbian Astronomical Journal, 184, 69

\bibitem[{{Bozzetto} {et~al.}(2012{\natexlab{b}}){Bozzetto}, {Filipovic},
  {Crawford}, {Payne}, {de Horta}, \& {Stupar}}]{2012RMxAA..48...41B}
{Bozzetto}, L.~M., {Filipovic}, M.~D., {Crawford}, E.~J., {et~al.}
  2012{\natexlab{b}}, \rmxaa, 48, 41

\bibitem[{{Bozzetto} {et~al.}(2012{\natexlab{c}}){Bozzetto}, {Filipovic},
  {Urosevic}, \& {Crawford}}]{2012SerAJ.185...25B}
{Bozzetto}, L.~M., {Filipovic}, M.~D., {Urosevic}, D., \& {Crawford}, E.~J.
  2012{\natexlab{c}}, Serbian Astronomical Journal, 185, 25

\bibitem[{{Bozzetto} {et~al.}(2014{\natexlab{a}}){Bozzetto}, {Filipovi{\'c}},
  {Uro{\v s}evi{\'c}}, {Kothes}, \& {Crawford}}]{2014MNRAS.440.3220B}
{Bozzetto}, L.~M., {Filipovi{\'c}}, M.~D., {Uro{\v s}evi{\'c}}, D., {Kothes},
  R., \& {Crawford}, E.~J. 2014{\natexlab{a}}, \mnras, 440, 3220

\bibitem[{{Bozzetto} {et~al.}(2012{\natexlab{d}}){Bozzetto}, {Filipovi{\'c}},
  {Crawford}, {Haberl}, {Sasaki}, {Uro{\v s}evi{\'c}}, {Pietsch}, {Payne}, {de
  Horta}, {Stupar}, {Tothill}, {Dickel}, {Chu}, \&
  {Gruendl}}]{2012MNRAS.420.2588B}
{Bozzetto}, L.~M., {Filipovi{\'c}}, M.~D., {Crawford}, E.~J., {et~al.}
  2012{\natexlab{d}}, \mnras, 420, 2588

\bibitem[{{Bozzetto} {et~al.}(2013){Bozzetto}, {Filipovi{\'c}}, {Crawford},
  {Sasaki}, {Maggi}, {Haberl}, {Uro{\v s}evi{\'c}}, {Payne}, {De Horta},
  {Stupar}, {Gruendl}, \& {Dickel}}]{2013MNRAS.432.2177B}
---. 2013, \mnras, 432, 2177

\bibitem[{{Bozzetto} {et~al.}(2014{\natexlab{b}}){Bozzetto}, {Kavanagh},
  {Maggi}, {Filipovi{\'c}}, {Stupar}, {Parker}, {Reid}, {Sasaki}, {Haberl},
  {Uro{\v s}evi{\'c}}, {Dickel}, {Sturm}, {Williams}, {Ehle}, {Gruendl}, {Chu},
  {Points}, \& {Crawford}}]{2014MNRAS.439.1110B}
{Bozzetto}, L.~M., {Kavanagh}, P.~J., {Maggi}, P., {et~al.} 2014{\natexlab{b}},
  \mnras, 439, 1110

\bibitem[{{Cajko} {et~al.}(2009){Cajko}, {Crawford}, \&
  {Filipovic}}]{2009SerAJ.179...55C}
{Cajko}, K.~O., {Crawford}, E.~J., \& {Filipovic}, M.~D. 2009, Serbian
  Astronomical Journal, 179, 55

\bibitem[{{Chen} {et~al.}(2006){Chen}, {Wang}, {Gotthelf}, {Jiang}, {Chu}, \&
  {Gruendl}}]{2006ApJ...651..237C}
{Chen}, Y., {Wang}, Q.~D., {Gotthelf}, E.~V., {et~al.} 2006, \apj, 651, 237

\bibitem[{{Crawford} {et~al.}(2008{\natexlab{a}}){Crawford}, {Filipovic}, {de
  Horta}, {Stootman}, \& {Payne}}]{2008SerAJ.177...61C}
{Crawford}, E.~J., {Filipovic}, M.~D., {de Horta}, A.~Y., {Stootman}, F.~H., \&
  {Payne}, J.~L. 2008{\natexlab{a}}, Serbian Astronomical Journal, 177, 61

\bibitem[{{Crawford} {et~al.}(2010){Crawford}, {Filipovi{\'c}}, {Haberl},
  {Pietsch}, {Payne}, \& {de Horta}}]{2010A&A...518A..35C}
{Crawford}, E.~J., {Filipovi{\'c}}, M.~D., {Haberl}, F., {et~al.} 2010, \aap,
  518, A35

\bibitem[{{Crawford} {et~al.}(2008{\natexlab{b}}){Crawford}, {Filipovic}, \&
  {Payne}}]{2008SerAJ.176...59C}
{Crawford}, E.~J., {Filipovic}, M.~D., \& {Payne}, J.~L. 2008{\natexlab{b}},
  Serbian Astronomical Journal, 176, 59

\bibitem[{{De Horta} {et~al.}(2012){De Horta}, {Filipovi{\'c}}, {Bozzetto},
  {Maggi}, {Haberl}, {Crawford}, {Sasaki}, {Uro{\v s}evi{\'c}}, {Pietsch},
  {Gruendl}, {Dickel}, {Tothill}, {Chu}, {Payne}, \&
  {Collier}}]{2012A&A...540A..25D}
{De Horta}, A.~Y., {Filipovi{\'c}}, M.~D., {Bozzetto}, L.~M., {et~al.} 2012,
  \aap, 540, A25

\bibitem[{{De Horta} {et~al.}(2014){De Horta}, {Sommer}, {Filipovi{\'c}},
  {O'Brien}, {Bozzetto}, {Collier}, {Wong}, {Crawford}, {Tothill}, {Maggi}, \&
  {Haberl}}]{2014arXiv1404.3823D}
{De Horta}, A.~Y., {Sommer}, E.~R., {Filipovi{\'c}}, M.~D., {et~al.} 2014,
  ArXiv e-prints

\bibitem[{{Filipovi{\'c}} {et~al.}(2008){Filipovi{\'c}}, {Haberl}, {Winkler},
  {Pietsch}, {Payne}, {Crawford}, {de Horta}, {Stootman}, \&
  {Reaser}}]{2008A&A...485...63F}
{Filipovi{\'c}}, M.~D., {Haberl}, F., {Winkler}, P.~F., {et~al.} 2008, \aap,
  485, 63

\bibitem[{{Foster} {et~al.}(2011){Foster}, {Smith}, {Ji}, \&
  {Brickhouse}}]{2011nlaw.confC...2F}
{Foster}, A.~R., {Smith}, R.~K., {Ji}, L., \& {Brickhouse}, N.~S. 2011, in 2010
  NASA Laboratory Astrophysics Workshop, C2

\bibitem[{{Freeman} {et~al.}(2001){Freeman}, {Doe}, \&
  {Siemiginowska}}]{2001SPIE.4477...76F}
{Freeman}, P., {Doe}, S., \& {Siemiginowska}, A. 2001, in Society of
  Photo-Optical Instrumentation Engineers (SPIE) Conference Series, Vol. 4477,
  Society of Photo-Optical Instrumentation Engineers (SPIE) Conference Series,
  ed. J.-L. {Starck} \& F.~D. {Murtagh}, 76--87

\bibitem[{{Freeman} {et~al.}(2002){Freeman}, {Kashyap}, {Rosner}, \&
  {Lamb}}]{2002ApJS..138..185F}
{Freeman}, P.~E., {Kashyap}, V., {Rosner}, R., \& {Lamb}, D.~Q. 2002, \apjs,
  138, 185

\bibitem[{{Fruscione} {et~al.}(2006){Fruscione}, {McDowell}, {Allen},
  {Brickhouse}, {Burke}, {Davis}, {Durham}, {Elvis}, {Galle}, {Harris},
  {Huenemoerder}, {Houck}, {Ishibashi}, {Karovska}, {Nicastro}, {Noble},
  {Nowak}, {Primini}, {Siemiginowska}, {Smith}, \&
  {Wise}}]{2006SPIE.6270E..60F}
{Fruscione}, A., {McDowell}, J.~C., {Allen}, G.~E., {et~al.} 2006, in Society
  of Photo-Optical Instrumentation Engineers (SPIE) Conference Series, Vol.
  6270, Society of Photo-Optical Instrumentation Engineers (SPIE) Conference
  Series

\bibitem[{{Ghavamian} {et~al.}(2007){Ghavamian}, {Laming}, \&
  {Rakowski}}]{2007ApJ...654L..69G}
{Ghavamian}, P., {Laming}, J.~M., \& {Rakowski}, C.~E. 2007, \apjl, 654, L69

\bibitem[{{Gooch}(1996)}]{1996ASPC..101...80G}
{Gooch}, R. 1996, in Astronomical Society of the Pacific Conference Series,
  Vol. 101, Astronomical Data Analysis Software and Systems V, ed. G.~H.
  {Jacoby} \& J.~{Barnes}, 80

\bibitem[{{Green}(2009{\natexlab{a}})}]{2009BASI...37...45G}
{Green}, D.~A. 2009{\natexlab{a}}, Bulletin of the Astronomical Society of
  India, 37, 45

\bibitem[{{Green}(2009{\natexlab{b}})}]{GREEN-CATA}
---. 2009{\natexlab{b}}, A Catalogue of Galactic Supernova Remnants
  (http://www.mrao.cam.ac.uk/surveys/snrs/)

\bibitem[{{Grondin} {et~al.}(2012){Grondin}, {Sasaki}, {Haberl}, {Pietsch},
  {Crawford}, {Filipovi{\'c}}, {Bozzetto}, {Points}, \&
  {Smith}}]{2012A&A...539A..15G}
{Grondin}, M.-H., {Sasaki}, M., {Haberl}, F., {et~al.} 2012, \aap, 539, A15

\bibitem[{{Haberl}(2004)}]{2004AdSpR..33..638H}
{Haberl}, F. 2004, Advances in Space Research, 33, 638

\bibitem[{{Haberl} {et~al.}(2000){Haberl}, {Filipovi{\'c}}, {Pietsch}, \&
  {Kahabka}}]{2000A&AS..142...41H}
{Haberl}, F., {Filipovi{\'c}}, M.~D., {Pietsch}, W., \& {Kahabka}, P. 2000,
  \aaps, 142, 41

\bibitem[{{Haberl} \& {Pietsch}(2008)}]{2008A&A...484..451H}
{Haberl}, F., \& {Pietsch}, W. 2008, \aap, 484, 451

\bibitem[{{Haberl} {et~al.}(2012{\natexlab{a}}){Haberl}, {Sturm},
  {Filipovi{\'c}}, {Pietsch}, \& {Crawford}}]{2012A&A...537L...1H}
{Haberl}, F., {Sturm}, R., {Filipovi{\'c}}, M.~D., {Pietsch}, W., \&
  {Crawford}, E.~J. 2012{\natexlab{a}}, \aap, 537, L1

\bibitem[{{Haberl} {et~al.}(2012{\natexlab{b}}){Haberl}, {Filipovi{\'c}},
  {Bozzetto}, {Crawford}, {Points}, {Pietsch}, {De Horta}, {Tothill}, {Payne},
  \& {Sasaki}}]{2012A&A...543A.154H}
{Haberl}, F., {Filipovi{\'c}}, M.~D., {Bozzetto}, L.~M., {et~al.}
  2012{\natexlab{b}}, \aap, 543, A154

\bibitem[{{Haberl} {et~al.}(2012{\natexlab{c}}){Haberl}, {Sturm}, {Ballet},
  {Bomans}, {Buckley}, {Coe}, {Corbet}, {Ehle}, {Filipovic}, {Gilfanov},
  {Hatzidimitriou}, {La Palombara}, {Mereghetti}, {Pietsch}, {Snowden}, \&
  {Tiengo}}]{2012A&A...545A.128H}
{Haberl}, F., {Sturm}, R., {Ballet}, J., {et~al.} 2012{\natexlab{c}}, \aap,
  545, A128

\bibitem[{{Hamilton} {et~al.}(1983){Hamilton}, {Sarazin}, \&
  {Chevalier}}]{1983ApJS...51..115H}
{Hamilton}, A.~J.~S., {Sarazin}, C.~L., \& {Chevalier}, R.~A. 1983, \apjs, 51,
  115

\bibitem[{{Hilditch} {et~al.}(2005){Hilditch}, {Howarth}, \&
  {Harries}}]{2005MNRAS.357..304H}
{Hilditch}, R.~W., {Howarth}, I.~D., \& {Harries}, T.~J. 2005, \mnras, 357, 304

\bibitem[{{Ishibashi} \& {Courvoisier}(2010)}]{2010A&A...512A..58I}
{Ishibashi}, W., \& {Courvoisier}, T.~J.-L. 2010, \aap, 512, A58

\bibitem[{{Johanson} \& {Kerton}(2009)}]{2009AJ....138.1615J}
{Johanson}, A.~K., \& {Kerton}, C.~R. 2009, \aj, 138, 1615

\bibitem[{{Kahabka} {et~al.}(1999){Kahabka}, {Pietsch}, {Filipovi{\'c} }, \&
  {Haberl}}]{1999A&AS..136...81K}
{Kahabka}, P., {Pietsch}, W., {Filipovi{\'c} }, M.~D., \& {Haberl}, F. 1999,
  \aaps, 136, 81

\bibitem[{{Kavanagh} {et~al.}(2013){Kavanagh}, {Sasaki}, {Points},
  {Filipovi{\'c}}, {Maggi}, {Bozzetto}, {Crawford}, {Haberl}, \&
  {Pietsch}}]{2013A&A...549A..99K}
{Kavanagh}, P.~J., {Sasaki}, M., {Points}, S.~D., {et~al.} 2013, \aap, 549, A99

\bibitem[{{Liedahl} {et~al.}(1995){Liedahl}, {Osterheld}, \&
  {Goldstein}}]{1995ApJ...438L.115L}
{Liedahl}, D.~A., {Osterheld}, A.~L., \& {Goldstein}, W.~H. 1995, \apjl, 438,
  L115

\bibitem[{{Maccacaro} {et~al.}(1988){Maccacaro}, {Gioia}, {Wolter}, {Zamorani},
  \& {Stocke}}]{1988ApJ...326..680M}
{Maccacaro}, T., {Gioia}, I.~M., {Wolter}, A., {Zamorani}, G., \& {Stocke},
  J.~T. 1988, \apj, 326, 680

\bibitem[{{Maggi} {et~al.}(2012){Maggi}, {Haberl}, {Bozzetto}, {Filipovi{\'c}},
  {Points}, {Chu}, {Sasaki}, {Pietsch}, {Gruendl}, {Dickel}, {Smith}, {Sturm},
  {Crawford}, \& {De Horta}}]{2012A&A...546A.109M}
{Maggi}, P., {Haberl}, F., {Bozzetto}, L.~M., {et~al.} 2012, \aap, 546, A109

\bibitem[{{Mao} {et~al.}(2008){Mao}, {Gaensler}, {Stanimirovi{\'c}},
  {Haverkorn}, {McClure-Griffiths}, {Staveley-Smith}, \&
  {Dickey}}]{2008ApJ...688.1029M}
{Mao}, S.~A., {Gaensler}, B.~M., {Stanimirovi{\'c}}, S., {et~al.} 2008, \apj,
  688, 1029

\bibitem[{{Mazzotta} {et~al.}(1998){Mazzotta}, {Mazzitelli}, {Colafrancesco},
  \& {Vittorio}}]{1998A&AS..133..403M}
{Mazzotta}, P., {Mazzitelli}, G., {Colafrancesco}, S., \& {Vittorio}, N. 1998,
  \aaps, 133, 403

\bibitem[{{McEntaffer} {et~al.}(2012){McEntaffer}, {Brantseg}, \&
  {Presley}}]{2012ApJ...756...17M}
{McEntaffer}, R.~L., {Brantseg}, T., \& {Presley}, M. 2012, \apj, 756, 17

\bibitem[{{M\"uller} {et~al.}(2003){M\"uller}, {Staveley-Smith}, {Zealey}, \&
  {Stanimirovi{\'c}}}]{2003MNRAS.339..105M}
{M\"uller}, E., {Staveley-Smith}, L., {Zealey}, W., \& {Stanimirovi{\'c}}, S.
  2003, \mnras, 339, 105

\bibitem[{{Owen} {et~al.}(2011){Owen}, {Filipovi{\'c}}, {Ballet}, {Haberl},
  {Crawford}, {Payne}, {Sturm}, {Pietsch}, {Mereghetti}, {Ehle}, {Tiengo},
  {Coe}, {Hatzidimitriou}, \& {Buckley}}]{2011A&A...530A.132O}
{Owen}, R.~A., {Filipovi{\'c}}, M.~D., {Ballet}, J., {et~al.} 2011, \aap, 530,
  A132

\bibitem[{{Payne} {et~al.}(2007){Payne}, {White}, {Filipovi{\'c}}, \&
  {Pannuti}}]{2007MNRAS.376.1793P}
{Payne}, J.~L., {White}, G.~L., {Filipovi{\'c}}, M.~D., \& {Pannuti}, T.~G.
  2007, \mnras, 376, 1793

\bibitem[{{Petre} {et~al.}(2007){Petre}, {Hwang}, {Holt}, {Safi-Harb}, \&
  {Williams}}]{2007ApJ...662..988P}
{Petre}, R., {Hwang}, U., {Holt}, S.~S., {Safi-Harb}, S., \& {Williams}, R.~M.
  2007, \apj, 662, 988

\bibitem[{{Reynolds}(1998)}]{1998ApJ...493..375R}
{Reynolds}, S.~P. 1998, \apj, 493, 375

\bibitem[{{Reynolds} \& {Keohane}(1999)}]{1999ApJ...525..368R}
{Reynolds}, S.~P., \& {Keohane}, J.~W. 1999, \apj, 525, 368

\bibitem[{{Russell} \& {Dopita}(1992)}]{1992ApJ...384..508R}
{Russell}, S.~C., \& {Dopita}, M.~A. 1992, \apj, 384, 508

\bibitem[{{Sault} {et~al.}(1995){Sault}, {Teuben}, \&
  {Wright}}]{1995ASPC...77..433S}
{Sault}, R.~J., {Teuben}, P.~J., \& {Wright}, M.~C.~H. 1995, in Astronomical
  Society of the Pacific Conference Series, Vol.~77, Astronomical Data Analysis
  Software and Systems IV, ed. R.~A. {Shaw}, H.~E. {Payne}, \& J.~J.~E.
  {Hayes}, 433

\bibitem[{{Sault} \& {Wieringa}(1994)}]{1994A&AS..108..585S}
{Sault}, R.~J., \& {Wieringa}, M.~H. 1994, \aaps, 108, 585

\bibitem[{{Smith} {et~al.}(2001){Smith}, {Brickhouse}, {Liedahl}, \&
  {Raymond}}]{2001ApJ...556L..91S}
{Smith}, R.~K., {Brickhouse}, N.~S., {Liedahl}, D.~A., \& {Raymond}, J.~C.
  2001, \apjl, 556, L91

\bibitem[{{Str{\"u}der} {et~al.}(2001){Str{\"u}der}, {Briel}, {Dennerl},
  {Hartmann}, {Kendziorra}, {Meidinger}, {Pfeffermann}, {Reppin}, {Aschenbach},
  {Bornemann}, {Br{\"a}uninger}, {Burkert}, {Elender}, {Freyberg}, {Haberl},
  {Hartner}, {Heuschmann}, {Hippmann}, {Kastelic}, {Kemmer}, {Kettenring},
  {Kink}, {Krause}, {M{\"u}ller}, {Oppitz}, {Pietsch}, {Popp}, {Predehl},
  {Read}, {Stephan}, {St{\"o}tter}, {Tr{\"u}mper}, {Holl}, {Kemmer}, {Soltau},
  {St{\"o}tter}, {Weber}, {Weichert}, {von Zanthier}, {Carathanassis}, {Lutz},
  {Richter}, {Solc}, {B{\"o}ttcher}, {Kuster}, {Staubert}, {Abbey}, {Holland},
  {Turner}, {Balasini}, {Bignami}, {La Palombara}, {Villa}, {Buttler},
  {Gianini}, {Lain{\'e}}, {Lumb}, \& {Dhez}}]{2001A&A...365L..18S}
{Str{\"u}der}, L., {Briel}, U., {Dennerl}, K., {et~al.} 2001, \aap, 365, L18

\bibitem[{{Sturm} {et~al.}(2013){Sturm}, {Haberl}, {Pietsch}, {Ballet},
  {Hatzidimitriou}, {Buckley}, {Coe}, {Ehle}, {Filipovi{\'c}}, {La Palombara},
  \& {Tiengo}}]{2013A&A...558A...3S}
{Sturm}, R., {Haberl}, F., {Pietsch}, W., {et~al.} 2013, \aap, 558, A3

\bibitem[{{Telezhinsky} {et~al.}(2012){Telezhinsky}, {Dwarkadas}, \&
  {Pohl}}]{2012APh....35..300T}
{Telezhinsky}, I., {Dwarkadas}, V.~V., \& {Pohl}, M. 2012, Astroparticle
  Physics, 35, 300

\bibitem[{{Udalski} {et~al.}(1998){Udalski}, {Szymanski}, {Kubiak},
  {Pietrzynski}, {Wozniak}, \& {Zebrun}}]{1998AcA....48..147U}
{Udalski}, A., {Szymanski}, M., {Kubiak}, M., {et~al.} 1998, \actaa, 48, 147

\bibitem[{{van der Heyden} {et~al.}(2004){van der Heyden}, {Bleeker}, \&
  {Kaastra}}]{2004A&A...421.1031V}
{van der Heyden}, K.~J., {Bleeker}, J.~A.~M., \& {Kaastra}, J.~S. 2004, \aap,
  421, 1031

\bibitem[{{Williams} {et~al.}(2005){Williams}, {Chu}, {Dickel}, {Gruendl},
  {Seward}, {Guerrero}, \& {Hobbs}}]{2005ApJ...628..704W}
{Williams}, R.~M., {Chu}, Y.-H., {Dickel}, J.~R., {et~al.} 2005, \apj, 628, 704

\bibitem[{{Wilms} {et~al.}(2000){Wilms}, {Allen}, \&
  {McCray}}]{2000ApJ...542..914W}
{Wilms}, J., {Allen}, A., \& {McCray}, R. 2000, \apj, 542, 914

\end{thebibliography}

\end{document}